\documentclass[twocolumn,showpacs,preprintnumbers,amsmath,amssymb]{revtex4}
\usepackage{graphicx}% Include figure files
\usepackage{dcolumn}% Align table columns on decimal point
\usepackage{bm}% bold math

\begin{document}

\preprint{APS/123-QED}

\title{Entrapping a polymer chain in a light well under a good solvent condition}

\author{Masatoshi Ichikawa}
\author{Kenichi Yoshikawa}
 \email{yoshikaw@scphys.kyoto-u.ac.jp}
\affiliation{Department of Physics, Graduate School of Science, Kyoto University \& CREST, Kyoto 606-8502, Japan}

\author{Yukiko Matsuzawa}
\affiliation{Department of Home Economics, Otsuma Women's University, 12 Sanban-cho, Chiyoda-ku, Tokyo 102-8357, Japan}

\date{\today}

\begin{abstract}
During the last decade, the laser trapping technique has been actively applied to elucidate the property of macromolecules, DNA, RNA, cytoskeleton fibres, etc. Due to the inherent difficulty in the direct trapping of single molecules at approximately 300 K, most of the current studies on laser trapping utilize the experimental technique to grasp microbeads that are attached to a single macromolecule, instead of using direct trapping. A few studies have demonstrated the applicability of laser trapping for highly packed compact DNA under a `poor' solvent condition without any beads. The present study achieves laser trapping on a single DNA molecule under a `good' solvent condition without any microbeads. The time-dependent change of the conformation accompanied by ON/OFF laser irradiation was measured and analysed in terms of the entrapment and release of a coiled polymer around a micrometer-sized potential. 
\end{abstract}

\pacs{36.20.-r,87.80.Cc,87.83.+a}% PACS, the Physics and Astronomy
                             % Classification Scheme.
\keywords{Optical tweezers, Laser trap, DNA, Rouse time}%Use showkeys class option if keyword
                              %display desired
\maketitle

%\section{Introduction}

Recently, studies on mechanical manipulations for single macromolecules have attracted great interest. These studies are expected to contribute to the development of polymer physics \cite{Gennes-SCPP,Chu-S,Perkins-S2,Doi-TPD,Perkins-S1,Gennes-JCP,Perkins-S3,Schroeder-S} and to initiate a new field in biophysics, e.g. force spectroscopy for RNA polymerization, DNA-histone interaction, etc., to elucidate novel insights on vital reactions \cite{Bustamante-N}. 

Until now, mechanical handling of a DNA molecule in water has been demonstrated using an alternating current electric field \cite{Ueda-PJ,Kaji-BJ}, a hydrodynamic force to stretch \cite{Schroeder-S} or collect \cite{Braun-PRL}, etc. Among the techniques, laser trapping or optical tweezing are frequently used \cite{Chu-S,Bustamante-N} because they can be applied to single DNA molecules by grasping a DNA-bound microsphere. A recent investigation demonstrated that a single DNA is pinched with 0.1 $\mu $m radii beads clustering \cite{Hirano-APL}. It has been reported that direct optical trapping for a single DNA chain is preferable such indirect trapping using microspheres in some terms. Folded compact DNA, such as supercoiled lambda phage DNA \cite{Chiu-JACS} and globular T4 phage DNA \cite{Katsura-NAR}, have been directly trapped and transported in water using optical tweezers. Through the active trials of optical trapping, it has become clear that stable trapping on an elongated coiled DNA is rather difficult and virtually impossible without the attachment of microbeads to them. Thus, the a posteriori results indicate that the optical gradient force affecting the DNA segments is insufficient to overcome thermal fluctuation during the bead pulling force spectroscopy experiments. In the present study, we report that an infrared focused laser stably traps an unfolded T4 DNA in a polymer solution (containing poly(ethylene glycol) (PEG)) with moderate optical power (several hundreds mW).

%\section{Materials and Method}

We used the T4 phage dsDNA (166 kbp, $\sim $56 $\mu $m, Nippon gene) stained with DAPI (4',6-diamidino-2-phenylindol, Wako Pure Chemical Industries, Ltd.) for the fluorescent microscopy. For microscopic observations, DNA with a base-pair concentration of 0.1 $\mu $M was enclosed in a glass cell along with Na phosphate buffer 7.8 mM, mercaptoethanol 1.5 \%(v/v) and poly(ethylene glycol) (MW avg. 6000 Da, Chameleon Reagent) 77 mg/ml. 
It has been well established that the presence of salt is necessary to condense a DNA molecule, and that this process is called the polymer-and-salt induced condensation, i.e. $\psi $-condensation \cite{Lerman-PNAS,Bloomfield-B,Minagawa-B,Yoshikawa-PD,Vasilevskaya-JCP,Bloomfield-COSB}. In the present study, we have adopted a low salt concentration well below the critical value of the $\psi $-condensation. 
From the transition point, it has been confirmed that the DNA coils collapse into folded globules in the presence of NaCl above 200 mM at such a PEG concentration \cite{Minagawa-B,Yoshikawa-PD,Vasilevskaya-JCP}. This implies that the DNA used in this study is situated in a good solvent condition. Experimental observations and optical trapping were performed using a fluorescent microscope (TE-300, Nikon) with a large-aperture oil-immersion objective lens (Plan Fluor $\times $100, NA = 1.30, Nikon). An Nd:YAG laser (Millennia IR, cw 1064 nm, TEM$_{00} $, Spectra Physics) for optical trapping was introduced into the objective lens using a dichroic mirror and was focused to a point approximately 1 $\mu $m in diameter on the observation field. The source laser power was 500 mW. Microscopic fluorescent images were detected by a high sensitivity SIT video camera and recorded on S-VHS videotapes.

%\section{Results}

\begin{figure}
\includegraphics{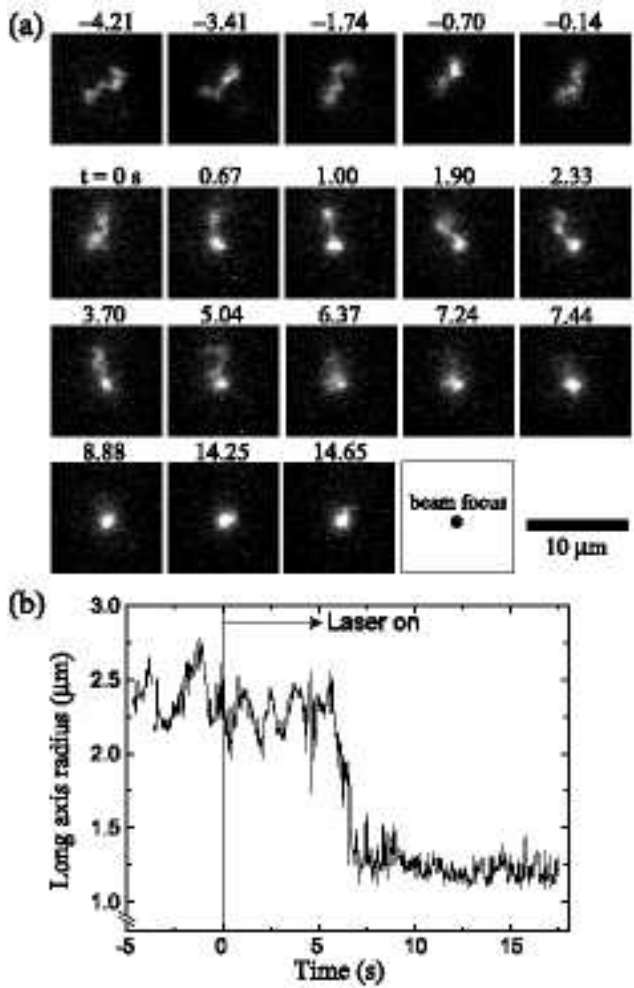}
\caption{\label{fig:Fig1.eps} A typical trapping process of coiled DNA. (a) Typical trapping process of a single T4 DNA. Time descriptions on each panel correspond to the time axis in the following graph. The trap laser is irradiated at time zero, and is focused on the centre of each picture in order to achieve the final schematic panel. (b) Time traces of the long axis radii of the DNA. }
\end{figure}

Figure 1 shows a typical result of the DNA-trapping experiment. The free DNA in the PEG solution as indicated in the upper panels of Fig. 1(a) exhibits intrachain and translational Brownian motion, indicating that the DNA adopts a random coil conformation in a good solvent condition. Figure 1(a) presents the conformational change after the start of the laser irradiation, focused close to the DNA, at time zero. The time traces of the DNA sizes and the focal brightness during this process are indicated in Fig. 1(b). The entire process or the effect of the laser illumination is described as follows. A certain part of the coil is first attracted towards the laser focus, and the remaining segments are then gradually pulled into focus (see supplementary movie). The remaining part of the coil retains its fluctuation before the entire coil was captured by the laser focus. A similar feature of the conformational change in single DNA has been encountered in the compaction in a poor solvent condition (salt containing PEG solution), where an almost linear folding process was observed after rather long induction period (time required to reach equilibrium) of several tens of minutes \cite{He-B}. Such a characteristic kinetic process of single-chain compaction has been interpreted in terms of nucleation and growth, inherent to the kinetics of first-order phase transitions. In order to clarify the intrinsic difference between the trapping of a single polymer in a good solvent condition and the DNA compaction in poor solvent condition, we have analysed the time-dependent experimental parameters for the release of chains.

%\section{Discussion}

The diffusion of the fluorescent centre of the object is plotted in Fig. 2(a). The results show that the DNA is definitely trapped on the laser focus, and that the effective potential can be calculated from the motion of the trapped DNA, as shown in Fig. 2(b). The solid line indicates an estimation of the effective trapping potential on a sphere with a radius of 0.7 $\mu$m \cite{Tlusty-PRL}. The mean square displacement (MSD) of the fluorescent center of the DNA shows that under a delay time of 0.3 s, the MSDs of trapped DNA are almost the same as those of the free ones. For a longer delay time, the MSDs of the trapped DNA, approximately 0.3 $\mu $m$^2$, reach the ceiling. These results indicate that the DNA pinned to the laser focus maintains the coiled state, without tight compaction. The fluorescent images of the trapped DNA also support the loose packing around the focus without tight compaction. The trapped DNA, as shown in Fig. 1, shrinks to approximately 1/50$^{th}$ of the visible volume from the free state. It has been established that DNA as a semi-flexible polyelectrolyte polymer exhibits a discrete coil-to-globule transition accompanied by a decrease in size from 1/1000 to 1/100000 \cite{Lerman-PNAS,Bloomfield-B,Minagawa-B,Yoshikawa-PD,Vasilevskaya-JCP,Bloomfield-COSB}. Thus, the volume decrease in our experiment is much smaller than the change in the folding transition.

\begin{figure}
\includegraphics{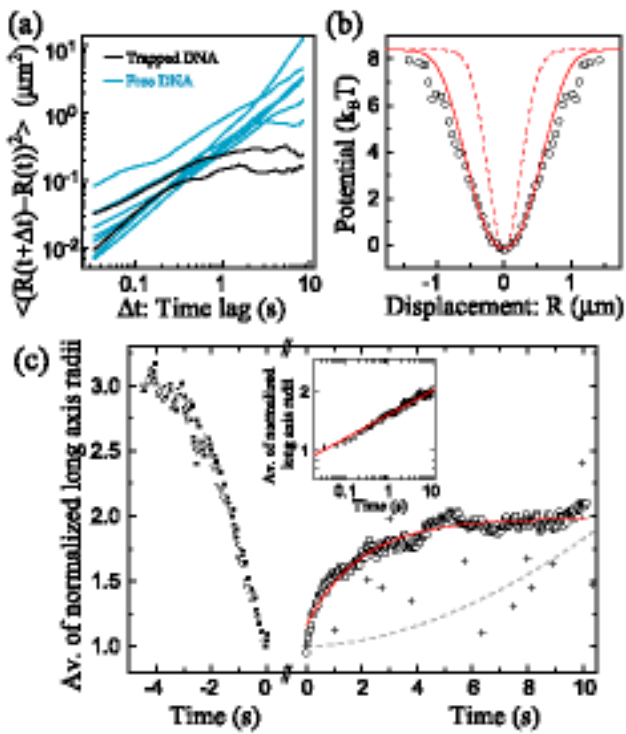}
\caption{\label{fig:Fig2.eps} Analyses of the motion of free and trapped DNA coils in the PEG solution. (a) Mean square displacements of trapped DNA and freely diffused DNA in logarithmic scale. (b) The effective trap potential deduced from the core motion of trapped DNA. The effective potential $ U(r) $ is obtained from the radial probability $ P(r)$: $ U(r) = -k_\mathrm{B} T \ln P(r)  + U_0 $. The centre of the coordinate is set on the arithmetic mean of the probability for each x-y axis. The fitting line (solid line) is the calculated effective potential of a particle with a radius of 0.7 $\mu $m in focused Gaussian light: radial intensity $ \propto \exp (-2r^2 / r_w^2) $, ($r_w = 0.5$ $\mu $m) (broken line). The depths of the lines are arbitrary. (c) The entrapping (22 av., solid circle) and releasing (27 av., open circle) process of DNA under 120 mg/ml PEG solution. The vertical axis is normalized on the average of the trapped DNA radii. Time zero of the trapping and swelling process are uniform at the time of completing engulfment (visual observation) and laser off, respectively. The exponential fitting line is $ 1.98 - 0.817 \exp (t/2.00) $. The broken line on the graph is the time development of the unfolding transition referred from \cite{Yoshikawa-JACS} ($l \sim t^{1.8}$). Logarithmic plots in the small window indicate the release process, fitted with $ \sim t^{0.125}$.}
\end{figure}

Figure 2(c) shows the time development of the long axis radii of the DNA swelling that started with the laser off (open circle) under a 120 mg/ml PEG concentration (at 2.26 mPa in viscosity). The graph shows additional evidences that the DNA is in the coiled state. In this process, the DNA exhibits a fast swelling at the initial stage, then slows down, and finally attains the elongated coil state within several seconds. This behaviour is quite different from the unfolding kinetics from the globule state to the coil state, as reported previously \cite{Yoshikawa-JACS}, where the unfolding is slow during the initial period and then accelerates (plotted as a broken grey line in Fig. 2(c)). The present behaviour as shown in the double logarithmic scale in Fig. 2(c), can be interpreted as a spring out phenomenon of a compressed coiled polymer as follows. The free energy of a compressed ideal polymer chain is denoted using the coil volume $ V $ as, 
\begin{equation}
F = \frac{3}{2} k_B T \frac{Nb^2}{V^{2/3}}.
\end{equation}
The chain pressure acting on the spherical wall is \cite{Grosberg-SPM}, 
\begin{equation}
p = - \frac{\partial F}{\partial V} \sim - k_B T \frac{Nb^2}{V^{5/3}}.
\end{equation}
Assuming the time development of the volume as $\partial V / \partial t \sim p $, (or ${\partial V}/{\partial t} \sim {\partial F}/{\partial V} $), the radial dilation speed is given by 
\begin{equation}
\frac{\mathrm{d} l}{\mathrm{d} t} \sim - k_B T \frac{Nb^2}{l^7},
\end{equation}
where the length of the axis of the coil is assumed to be $l \sim V^{1/3} $. Thus, we achieve $l \sim t^{\frac{1}{8}} $, which is in good correspondence with the experimental result $t^{0.125} $, as shown in Fig. 2(c). We have to be cautious regarding this `coincidence' because of the assumption that an ideal polymer does not reflect an experimental condition, and an experiment also has measurement errors. However, the coil invariably results in this type of dilation. Thus, we deal with the trapped DNA as a shrunk coil in order to analyses the DNA in an external potential.

Subsequently, we focus our attention on the rotational relaxation time of the coil shape in each free (open circle) and trapped (solid circle) state. Figure 3 shows the autocorrelations of a long axis vector of the DNA in each state. Although the orientation time of an end-to-end vector of the coil corresponds to the Rouse time, it is difficult to measure the exact magnitude of the vector from the fluorescent images. We use a long axis vector in place of an end-to-end vector to approximately estimate the orientation time. From the experimental results of the autocorrelations of the vector in trapped and free coils, it is observed that the rapid damping of the first frame in each condition is attributed to the fact that the long axis vector frequently changes its direction discontinuously, depending on the change in shape, differently from the end-to-end vector. In addition, since the vector angle is defined from 0 to $\pi$, a half full angle, during the process of quantifying the experimental 2D images, the vector angle sometimes flips from 0 to $\pi $ (or the inverse). Such mechanical problems, in this case, cannot be disassociated from the data except for the subjective correction. We adopt the raw datas to calculate an orientation time of the coil shape. From the fitting without the first lags, the relaxation times of the coils achieved as $ \tau_R= 3.1$ [s] (free) and $ \tau_R' = 0.17 $ [s] (trapped). 

\begin{figure}
\includegraphics{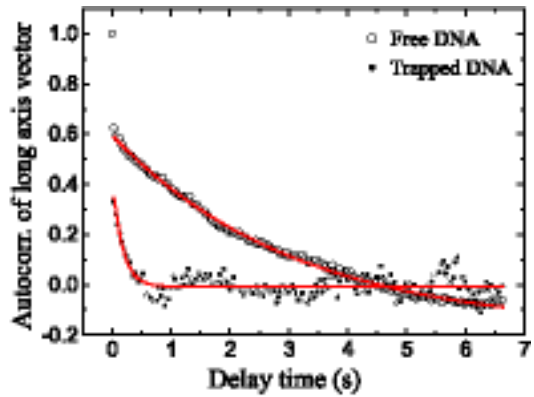}
\caption{\label{fig:Fig3.eps} The average of the autocorrelation of the long axis length of a fluctuating DNA in the trapped and free states. The long axis length is the longest line between the corner points of a fluorescent spot. The fitting functions are $ - 0.007 + 0.434 \exp (t/0.169) $ (trapped) and $ -0.181 + 0.778 \exp (t/3.11) $ (free), respectively. }
\end{figure}

Next, we compare the above results with the Rouse model. The Rouse time, $\tau_R $, is expressed as,
\begin{equation}
\tau_R = \frac{N^2 b^2 \zeta }{3 \pi^2 p^2 k_\mathrm{B} T} ,
\end{equation}
where $N$, $b$ and $\zeta$ are number of segments, the segment length and the viscous drag coefficient per segment, respectively. We will evaluate these variables from the experimental values. The viscous drag coefficient of the whole coil, $\zeta_0 $, is simply denoted as $ \zeta \simeq \zeta_0 / N $ neglecting the fluid effect. Einstein's relation is given as $ \zeta_0 = k_\mathrm{B} T / D_0 $, where $D_0$ is the diffusion coefficient of the coil. The variable $ D_0 $ is evaluated from the fitting data from 0 to 2 s as shown in Fig. 2(a) as $ D_0 = 0.081 $ [($\mu $m)$^2$/s] and $N b^2$ is $ 5.6 $ [($\mu $m)$^2$], where we assume $ b = 100 $ [nm] as the Kuhn length and $ N = 560 $ \cite{Comment-1}. Thus, under these experimental conditions, we obtain $ \tau_R = 2.3 $ [s] for $ p = 1 $. 

Wenczel et al. \cite{Wenczel-JCP} have derived the Rouse time in a strong harmonic potential: $ \bar{k} \mathbf{r}^2 /2 $. They achieved the analytical result that the orientation time shortens depending on the strength of the external potential as 
\begin{equation}
\tau_R' = \frac{N^2 b^2 \zeta }{3 \pi^2 p^2 k_\mathrm{B} T + N^2 b^2 \bar{k}} . 
\label{eq:Wenczel}
\end{equation}
The depth of the effective potential is measured to be $ \Delta U_{total} \sim $ 8 [$k_\mathrm{B} T$] as shown in Fig. 2(c); thus, the field strength for one segment is $  \Delta U_{segment} = \Delta U_{total}/N $. On the other hand, the Gauss type trapping potential is approximated to be the harmonic potential $ \bar{k} (0.4 [\mu \mathrm{m}])^2 /2 \simeq \Delta U_{segment} $. Thus, $ \bar{k} \simeq 100/N ~[{k_\mathrm{B} T}/{(\mu \mathrm{m})^2}] $ is obtained. Since the approximation that $\frac{N^2 b^2 \bar{k}}{3 \pi^2 p^2 k_\mathrm{B} T} \gg 1 $ is advisable for small $p$, eq. (\ref{eq:Wenczel}) immediately yields the following value of $ \tau_R $ without $ N^2 b^2 $ dependence 
\begin{equation}
\tau_R' = \frac{\frac{N^2 b^2 \zeta }{3 \pi^2 p^2 k_\mathrm{B} T}}{1+ \frac{N^2 b^2 \bar{k}}{3 \pi^2 p^2 k_\mathrm{B} T}} \sim \frac{\zeta}{\bar{k}} \simeq \frac{1}{100 D_0} \simeq 0.12 ~~[\mathrm{s}].
\end{equation}
Thus, the obtained result provided a good explanation of the experimental trend. In summary, the relaxation time of the Rouse model for a strong harmonic potential qualitatively reproduces the experimental results.

%\section{Conclusion}

In conclusion, we have demonstrated the optical trapping of a coiled DNA, and have experimentally measured the polymer compression and release processes. The present study provided the technique to compress a polymer and demonstrated the effect of an external potential on a polymer, which can be utilised for fundamental research. This method can also be applied to DNA nano-biotechnology, e.g. a micro-well plate as in the abstract, and to the selection and detection by DNA compression. The possibility of sequence separation by spatiotemporal gradient light is possible on the basis of an analogy with pulsed field electrophoresis.

\begin{acknowledgments}
The authors wish to thank Mr. T. Harada, Dr. H. Oana, Dr. S. M. Nomura and Dr. H. Mayama for their helpful technical assistance. This study drew from an experiment performed along with Mr. M. Murata. The authors also wish to thank Prof. C. Y. Shew and Dr. M. Tanaka for their fruitful discussions. Author M. I. is supported by the JSPS Research Fellowships for Young Scientists. The financial support provided by the Grant-in-Aid for the 21st Century COE `Center for Diversity and Universality in Physics' from the Ministry of Education, Culture, Sports, Science and Technology (MEXT) of Japan is gratefully acknowledged. 
\end{acknowledgments}


\begin{thebibliography}{99}
\bibitem{Comment-1}Though $N b^2$ is fixable from the experiment, we do not use it because the environment is obviously a good solvent. The hydrodynamic radius, $ R_H $, of the coiled DNA is evaluated from Fig. 2(b) as $ R_H = 1.85 $ [$\mu $m] and the radius of gyration, $ R_G $, is denoted using $ R_H $ as $ R_G = 1.56 R_H $ \cite{Oono-JCP}. In $\theta$ state, $ N b^2 \simeq 6 R_G^2 \simeq 50 $ [($\mu $m)$^2$].
\end{thebibliography}
\end{document}